\documentclass[preprint2]{emulateapj}

\newcommand{\reduceme}{\mbox{R\raisebox{-0.35ex}{E}D%
\hspace{-0.05em}\raisebox{0.85ex}{uc}\hspace{-0.90em}%
\raisebox{-.35ex}{{m}}\hspace{0.05em}E}}

\shorttitle{Carbon and Nitrogen in Early-type Galaxies}
\shortauthors{Toloba et~al.}

\begin{document}


\title{Carbon and Nitrogen Abundances in Early-type Galaxies}


\author{E. Toloba\altaffilmark{1}}
\author{P. S\'{a}nchez-Bl\'{a}zquez\altaffilmark{2}}
\author{J. Gorgas\altaffilmark{1}}
\author{B.~K. Gibson\altaffilmark{2}}
\altaffiltext{1}{Universidad Complutense de Madrid, 28040, Madrid, Spain}
\altaffiltext{2}{Centre for Astrophysics, University of Central Lancashire, Preston, PR1~2HE, UK}

\begin{abstract} For the first time, we undertake a systematic
examination of the nitrogen abundances for a sample of 35 early-type
galaxies spanning a range of masses and local environment.  The
nitrogen-sensitive molecular feature at 3360\AA\ has been employed in
conjunction with a suite of atomic- and molecular-sensitive indices to
provide unique and definitive constraints on the chemical content of
these systems.  By employing NH3360, we are now able to break the
carbon, nitrogen, and oxygen degeneracies inherent to the use of the
CN-index.  We demonstrate that the NH3360 feature shows little
dependency upon the velocity dispersion (our proxy for mass) of the
galaxies, contrary to what is seen for carbon- and magnesium-sensitive
indices. At face value, these results are at odds with conclusions
drawn previously using indices sensitive to both carbon and nitrogen,
such as cyanogen (CN). With the aid of stellar population models, we
find that the N/Fe ratios in these galaxies are consistent with being
mildly-enhanced with respect to the solar ratio. We also explore the
dependence of these findings upon environment, by analyzing the
co-added spectra of galaxies in the field and the Coma cluster.  We
confirm the previously found differences in carbon abundances between
galaxies in low- and high-density environments, while showing that
these differences do not seem to exist for nitrogen. We discuss the
implications of these findings for the derivation of the star
formation histories in early-type galaxies, and for the origin of
carbon and nitrogen, themselves.
\end{abstract}


\section{Introduction}

Uncovering the star formation histories of early-type galaxies has
proved to be a challenging endeavor for several decades now.  Perhaps
the most common approach to date has been via the comparison of
Lick/IDS spectral indices, each with varying degrees of sensitivity to
age  and metallicity \citep[][]{Wor94}.  Alas, the lack of indices
which  are \it purely \rm sensitive to age and/or metallicity has
limited the uniqueness of the claimed results.  Despite a rich
literature,  there remains fundamental disagreement on the magnitude
and even  veracity of important trends, including those linking age,
metallicity, and relative elemental abundances, with mass  and
environment.  A further limitation to this  technique is its
sensitivity to the youngest stellar component in a given galaxy, as
opposed to the more representative underlying population.

An alternative approach, and one employed successfully for our own
Galaxy, is to use our knowledge of galactic chemical evolution to
inform our derivation of star formation histories.  In massive
early-type galaxies \citet{Oconnel76} first noticed that Mg abundance
was enhanced relative to Fe; Other works confirmed this by comparing
absorption line indices with isochrone-based stellar population
models. Because Mg is predominantly  produced in Type~II supernovae
while Fe production is  dominated by Type~Ia supernovae,  this
enhancement is often interpreted as the result of very  short
timescales for the formation of stars in these galaxies.  Apart from
Mg, other elements (eg. C, N, Na) have been claimed to be enhanced in
large ellipticals (Worthey 1998\nocite{Wor98}; Burstein
2003\nocite{Burs03}; Kelson et al.\ 2006\nocite{Kel06},
S\'anchez-Bl\'azquez et al. 2003\nocite{SB03}, 2006a\nocite{SB06a}
(SB03 and SB06a hereafter); Schiavon 2007\nocite{Schiav07}; Graves et
al. 2008\nocite{Grav08}) and to be strongly correlated with  their
velocity dispersions \citep[although see][]{Kel06,Clem06} for an
alternate perspective regarding Mg and C).

Claims have been made for a strong N/Fe correlation with mass
\citep[SB06a,][]{Kel06,Schiav07,Grav08}. Nitrogen is produced during
hydrogen burning via the CNO and CN cycles, and it is created as both,
primary and secondary element. In primary nucleosynthesis N is
produced  at the same time as C and O and it is independent of
metallicity while in secondary production nitrogen is synthesized from
the carbon and  oxygen already present in the star, and its abundance
is therefore proportional to the heavy elements abundance. The
correlation of N/Fe with mass have been interpreted as a correlation
between N/Fe and  metallicity (through the mass-metallicity relation)
and has lead the authors above to conclude that most of the N produced
in ellipticals is of secondary origin.  To derive nitrogen abundances,
the CN-band at  4175\AA~ has typically been used; however,
disentangling the effects of C, N, and O using the CN-band  is
difficult \citep[][]{Burs03}. Direct observation of  the NH feature at
3360\AA~ offers a more direct, and robust  estimator as the NH-feature
is insensitive to C and O \citep[][]{Sned73, Norr02} and it is directly measuring N abundances \citep[][]{B&N82,T&L84}.  In addition,
the near-UV light is produced primarily by dwarf stars
\citep[][]{DC94}, while the optical cyanogen indices result from a
more complex mix of both dwarf and giant star light. This avoid the
variation in the stellar atmospheric abundances due to mixing during
the first dredge-up.

There are fewer than 15 early-type galaxies with published values of 
NH3360 in the literature \citep[][]{Pon98,Bou88,DC94}, due in large
part to the relative insensitivity of detectors in the near-UV.
We present here observations of the NH feature
in a sample of 35 early-type galaxies 
in the field, Virgo, and Coma clusters.
We show that previous claims of correlations 
between nitrogen and velocity dispersion are incorrect 
and that the NH3360 index-$\sigma$ relation is almost flat.

\section{Observations and Data Reduction}

Long slit spectroscopic data for 35 ellipticals were obtained at the
4.2m William Herschel Telescope at Roque de los Muchachos Observatory,
using the ISIS spectrograph. The instrumental configuration provided
and spectral coverage from 3140 to 4040\AA\ in the blue arm with an
spectral resolution of 2.3\AA\ (FWHM) and a typical sigmal to noise per \AA{} 
of 40. Our sample is a subset of those
presented in SB06a, allowing us to supplement our near-UV data with
the associated optical indices (3500$-$5250\AA). We measure the NH3360
index in our new data, and the \nocite{Ser05}Serven, Worthey \& Briley
(2005, SWB05 hereafter) optical indices in our SB06a optical spectra.

In SB03, we claimed that galaxies in the Coma cluster had much lower
CN and C4668 indices than those in lower-density environments. We were
unable then to conclude whether the differences were due exclusively
to differences in carbon or if nitrogen was also overabundant in
galaxies in lower-density environments.  With the aim of resolving
this issue, we selected our sample to include field, Virgo and Coma
ellipticals, spanning a range in velocity dispersions
(130$<$$\sigma$$<$330~km~s$^{-1}$).  Following the nomenclature of
SB03, we denote our Coma cluster sample as High-Density Environment
Galaxies (HDEG), while galaxies in the field and Virgo are denoted
Low-Density Environment Galaxies (LDEG).

We have followed the standard reduction procedure for long-slit
spectra, using \reduceme~\citep[][]{Car99}.  Galaxy spectra were
extracted within a central equivalent aperture of 4$^{\prime\prime}$
at the distance of NGC~6703 (corresponding to a physical aperture of
0.76~kpc).  We measured the NH3360 index using the definition of
\citet{DC94}\footnote{A case could be made to amend the Davidge \&
  Clark (1994) definition somewhat, in light of Carbon et~al. (1992;
  Fig~13).  Specifically, the NH3360 continua sit within the broad
  wings of the NH feature itself, reducing its optimal sensitivity to
  nitrogen. Redefining the red continuum to be 3480$-$3520\AA\ would
  mitigate this effect to some degree, but for consistency with the
  canonical definition, we have not pursued this approach.}  and the
CNO3862, CNO4175, and CO4685 indices, as defined by SWB05.  All
indices were measured at a dispersion of $\sigma$=200~km~s$^{-1}$ and
corrected for the effects of velocity dispersion broadening using a
combination of synthetic spectra from Bruzual \& Charlot (2003, BC03
hereafter) and Vazdekis et~al. (2008, V08 hereafter).

\section{Results}

\begin{figure}
\resizebox{0.45\textwidth}{!}{\includegraphics[angle=-90]{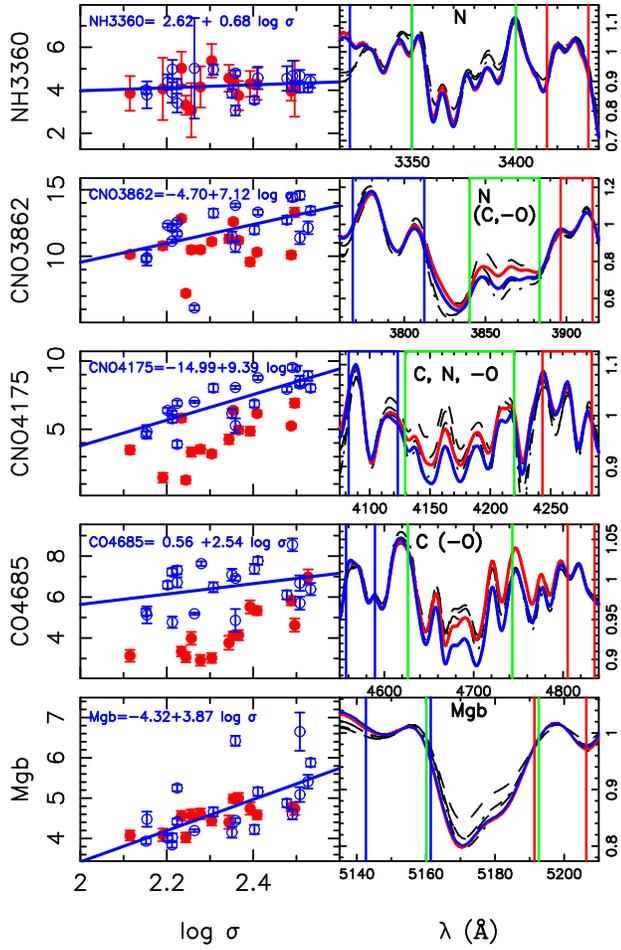}}
\caption{Left column: Line-strength indices versus velocity dispersion. Red 
solid circles correspond to HDEG, while blue open circles are LDEG.
Right column: Blue, green, and red boxes represent the bandpass of the 
index definition: In blue and red are the two-bands used to define the 
pseudocontinua. The flux in both, 
the data and the models have been normalized to these bands.
The central band where the feature is located is indicated in green.
Black lines represent stellar population models. Only models with solar 
(dashed lines)
and supersolar (dotted lines) metallicity 
are plotted; 6 and 15 Gyr models are plotted for each
metallicity.  The stacked spectra for LDEG are plotted in blue, 
while the HDEG are represented in red. Dominant species and other important 
species (within brackets) contributing to the indices defined by SWB05 are shown 
in the inset of the panels. A - symbol indicates that an increase in the abundance of that element makes the index weaker.
\label{figurel}}
\end{figure}

Figure~\ref{figurel} shows the co-added spectra in different
wavelength regions of LDEG (blue) and HDEG (red), with velocity
dispersions in the range 150$<$$\sigma$$<$250~km~s$^{-1}$.  Before
being added, the spectra were shifted to the same radial velocity and
broadened to the maximum $\sigma$ of all spectra (250~km~s$^{-1}$).
The spectral regions correspond to the bands of four different
indices: NH3360, CNO3862, CNO4175, and CO4685.

Overplotted in these diagrams are the models by BC03 for the NH3360
and the models of V08 for the remainder of the indices. We could not
use V08 for the comparison of the NH index because the spectra start
at 3500~\AA.  However, we prefer to use it for the other regions of
the spectra as the stellar library used in the BC03 models suffers
from wavelength calibration problems. Nevertheless, we have checked
and ensured that the results obtained are independent of the models
employed.  All synthetic spectra were degraded to the same broadened
resolution as the galactic spectra.  We have plotted four models in
each figure, with ages $\sim$6 and $15$~Gyr, and metallicities solar
and supersolar ([M/H]=+0.4 for BC03 and [M/H]=+0.2 for V08).

Comparing the empirical spectra with the models in Fig.~\ref{figurel}
and taking into account the element dependence of the indices (see
inset of the figure) we can see: NH3360 is slightly overabundant with
respect to the scaled-solar models; CNO3862 is compatible with the
scaled-solar models at lower metallicities (for older ages) or higher
metallicities (for younger ages); CNO4175 and CO4682 are overabundant
in the field galaxies, but not for the cluster galaxies.  In consort,
these data require nitrogen-based indices be overabundant with respect
to solar, carbon-based indices be overabundant with respect to solar
only for LDEG, and oxygen be similarly overabundant.

\subsection{The index-velocity dispersion relations}

In Fig.~\ref{figurel} we present the behavior of the four indices
analyzed, in addition to Mgb, as a function of the velocity dispersion
of the galaxies.  We distinguish between three different groups of
plots, depending on the chemical element dominating the intensity of
the index: N-based indices (CO3862 and NH3360), C-based indices
(CNO4175 and CO4685), and Mgb.  Indices dependent on C show a strong
dependency upon velocity dispersion, in a similar way to Mgb. However,
the trends between the N-sensitive indices and velocity dispersion are
much flatter; indeed, they are even flatter than that found for
$<$Fe$>$ (see SB03).

\begin{figure}
\resizebox{0.5\textwidth}{!}{\includegraphics[angle=-90]{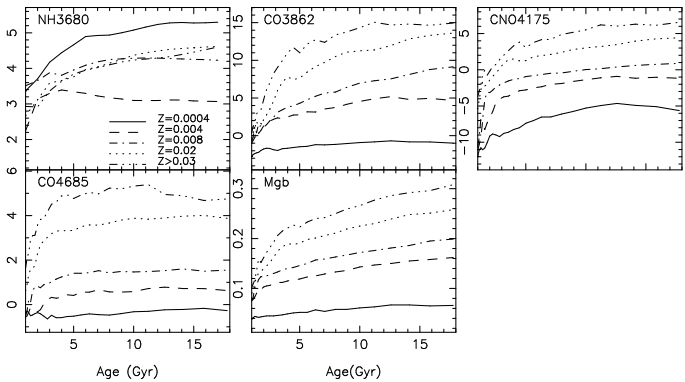}}
\caption{Variations of the analyzed indices with age at different
metallicities as indicated in the inset. BC03 models are used for the NH3360
while V08 models are employed for the remainder of the 
plots.\label{sensitivity}}
\end{figure}

Differences in the slope of the index-$\sigma$ diagram could appear
due to a different sensitivity of the indices to age and/or
metallicity and do not necessarily reflect differences in the chemical
abundance ratios.  Figure~\ref{sensitivity} shows the variation of the
indices with age for different metallicities. It can be seen that each
of the SWB05 indices show a strong dependence on metallicity,
particularly CO4685. The Lick/IDS index Mgb also shows a strong
metallicity dependence. On the other hand, for metallicities relevant
for elliptical galaxies ($Z>0.004$), NH3360 is essentially constant
with age and overall metallicity.  This behavior allows NH3360 to be
used successfully to derive nitrogen abundances in integrated spectra,
with \citet{Bou88} confirming its strong sensitivity to nitrogen
abundance .

Can we explain the relations between the other indices and $\sigma$ as
the consequence only of a relation between the overall metallicity
and/or age with $\sigma$?  To answer this question we have
parametrized the models of BC03 and V08 as a function of age and
metallicity, in the age and metallicity restricted to the range
covered by our sample, $\sim$4-$\sim$17 Gyrs and -0.70$<$[M/H]$<$0.40.
If we allow for the same age variation with $\sigma$ (from 5-12 Gyr in
the $\sigma$ range 125-350 kms$^{-1}$) for all the indices, we need
$\sim$0.13 dex more metallicity variation to explain the CNO4175- and
CO4685-$\sigma$ relations -- the most C-sensitive indices -- while and
extra $\sim$ 0.3 dex variation in metallicity is needed to explain the
relation of the Mgb-$\sigma$ slope. At a face value, this would be
indicating that [C/M], [Mg/M] and [Mg/C] increase with $\sigma$.  The
conclusions, however, remain very speculative until the oxygen
abundance and its variation with $\sigma$ is calculated in these
galaxies.  In addition, the indices strongly dependent on C show a
systematic offset between LDEG and HDEG, with the indices for HDEG
systematically weaker than those for LDEG.  \footnote{This behavior is
  also visible in the near-IR C-sensitive indices
  (M\'armol-Queralt\'o, in preparation).}

\section{Discussion}

We present here the NH3360 indices for a sample of 35 early-type
galaxies.  Unlike the ambiguities which exist when inferring nitrogen
abundances from cyanogen indices in the optical (due to contamination
from carbon and oxygen in the bands and associated continua), the
near-UV molecular feature of NH isolates the nitrogen abundance.  The
main conclusions of this work are: (1) a flat relation exists between
the NH3360 index and the velocity dispersion of the galaxies, contrary
to what has been previously claimed by other authors from their
analysis of the more ambiguous CN-index; (2) N is independent of local
environment, whereas C appears stronger for LDEG relative to HDEG. In
this sense, carbon and nitrogen production appear to be essentially
decoupled from one another.

Our results can be useful for constraining the origin of C and N.
Nitrogen is produced in both intermediate-mass and massive stars, and
it can have a primary and secondary origin
\citep[][]{Carig05,Chiapp06}.  The abundance ratio of a secondary
element is predicted to increase with metallicity.  If elliptical
galaxies show a mass-metallicity relation, as it has been claimed by
many authors \citep[eg.,][]{KA98}, then we should expect the secondary
N to increase with the mass of the galaxies, as suggested in previous
works based upon cyanogen-inferred nitrogen abundance determinations
\citep[eg.,][]{Kel06, Schiav07}. However, we have shown here that when
a more ``pure'' nitrogen-sensitive indicator is used, there is no
relation between the nitrogen abundance and the mass of the
galaxies. This lack of correlation may be indicating primary
production of nitrogen.

Nucleosynthesis studies predict that primary N can be produced both in
the third dredge-up along the Asymptotic Giant Branch (AGB) phase, if
nuclear burning at the base of the convective envelope is efficient
\citep[hot-bottom burning]{RV81}, and recent yields, including
rotationally-induced mixing, also predict significant production of N
in massive stars \citep{MM02}. \citet{Chiapp03, Chiapp06} has shown
that the inclusion of these yields in a Galactic chemical evolution
model of the Milky Way predicts an N/O gradient consistent with
observations.  Having said that, this primary nitrogen production
pathway is apparently only significant at low metallicities and is not
predicted to be particularly relevant for systems such as elliptical
galaxies that can reach supersolar metallicities very rapidly.  In a
system more similar to elliptical galaxies -- that of the bulge of our
galaxy -- \citet{Ballero07} \citep[see also][]{Matt86} need to include
primary N from massive stars of {\it all} metallicities (and all
masses), in order too explain the data from planetary nebulae.

In elliptical galaxies, the offsets between LDEG and HDEG favor
massive stars as the main contributors to primary nitrogen, because
for the elements produced by intermediate mass stars, no offsets are
detected. However, if that is the case, then we would expect a similar
correlation between N and $\sigma$, as seen for Mg and $\sigma$.  As
we have shown though, the N--$\sigma$ correlation appears flat from
our new data and analysis.  A possible solution is that the
contribution from {\it both} massive and intermediate mass stars is
shaping the N--$\sigma$ relation in ellipticals.  The delay in the N
release from intermediate mass (4$<$M $<$8 M$_{\odot}$) stars is
$\sim$100~Myr, so even in a relatively short star formation burst,
some contribution from these stars is expected.  It is believed that
the star formation in ellipticals is both more efficient and of a
shorter timescale with increasing galactic mass.  If this is the case,
N would be produced increasingly more prominently by massive stars in
the most massive galaxies, while the less-massive galaxies with their
more protracted star formation history would have an additional
contribution from intermediate mass stars.  As mentioned, the delay in
the release of N from intermediate mass stars is $\sim$100~Myrs (the
lifetime of 5$-$6~M$_\odot$ solar metallicity stars); intermediate
mass stars do not release significant amounts of Mg at high
metallicities \footnote{note that significant contribution from AGB
  stars to the heavy isotopes $^{25}$Mg and $^{26}$Mg is predicted at
  metallicities lower than [Fe/H]$<-0.5$ \citep{Fenn03}}, which would
explain the different behavior of the Mg$-$$\sigma$ relation compared
with the N-$\sigma$ relation.

Carbon is another controversial element.  While some authors argue
that it is mainly produced in intermediate mass stars
\citep[][]{Chiapp03} (with masses between 1$-$4~M$_{\odot}$) others
suggest that massive stars are the main contributors
\citep[eg.,][]{Carig05}.  The offset between galaxies in different
environments that is not seen in other elements produced mainly in
massive stars, such as Mg, favors the suggestion that C is produced
mainly in lower mass stars.  However, if low- and intermediate-mass
stars {\it are} the main producers of C, it is difficult to explain
the strong correlation between C and $\sigma$ and the overabundance of
C/Fe with respect to the Sun.  One possibility is that C is produced
in both, massive and lower mass stars in comparable proportions. 
The significant slope in the
relation would be the consequence of C production in massive stars
while the offset would come from a different production in lower mass
stars.  This latter C release would increase the C budget in LDEG
while in the HDEG, star formation ceases before this can happen.  

Of course these are very simplified scenarios.  The relative
importance of the different N-sites and the origin of C in elliptical
galaxies will be explored in a future paper using our {\tt GEtool}
chemical evolution code (eg., Hughes et~al. 2008).

In summary, we report the first determination of ``pure'' nitrogen
abundances for a sample of early-type galaxies and compare these with
stellar population models.  Contrary to what has been claimed
previously, the nitrogen abundances do not vary with the velocity
dispersion of the galaxies.  The environmental differences found for C
and N abundances impose an important constraint on chemical evolution
and galaxy formation models.  It will be critical to independently
constrain the oxygen abundance and its dependence upon galaxy mass, in
part because of the obvious synergies between CNO in low- and
intermediate-mass stars, and in part because oxygen is simply the most
dominant component ($\sim$50\%) of global metallicity.

\acknowledgments
We thank Francesca Matteucci and Reynier Peletier for several very
useful discussions. The expert guidance of the referee, Judy Cohen, is gratefully acknowledged. This work has been partially supported by the
Spanish research project AYA2007-67752-C03-03, the Marie Curie Intra-European
Fellowship scheme within the 6th European Community Framework
Programme (PSB), UK's Science \& Technology Facilities Council (BKG).
E.T. acknowledges the support of the University of Central
Lancashire's Livesey Award scheme (PSB).

\bibliographystyle{aa}
\bibliography{references}{}

\clearpage

\end{document}